# Electronic Noise Spectroscopy of Quasi-2D van der Waals Antiferromagnetic Semiconductors

Subhajit Ghosh[1,2], Zahra Ebrahim Nataj[1,2], Fariborz Kargar[1,2,×] and Alexander A. Balandin[1,2,×]

[1]Department of Materials Science and Engineering, University of California, Los Angeles, California 90095 USA

[2]California NanoSystems Institute, University of California, Los Angeles, California 90095 USA

**Abstract**

We investigated low-frequency current fluctuations, *i.e.* electronic noise, in FePS$_3$ van der Waals, layered antiferromagnetic semiconductor. The noise measurements have been used as *noise spectroscopy* for advanced materials characterization of the charge carrier dynamics affected by spin ordering and trapping states. Owing to the high resistivity of the material, we conducted measurements on vertical device configuration. The measured noise spectra reveal pronounced Lorentzian peaks of two different origins. One peak is observed only near the Néel temperature and it is attributed to the corresponding magnetic phase transition. The second Lorentzian peak, visible in the entire measured temperature range, has the characteristics of the trap-assisted generation-recombination processes similar to those in conventional semiconductors but shows a clear effect of the spin order reconfiguration near the Néel temperature. The obtained results contribute to understanding the electron and spin dynamics in this type of antiferromagnetic semiconductors and demonstrate the potential of electronic noise spectroscopy for advanced materials characterization.

---

[×] Corresponding authors: f.kargar@ucla.edu (F.K.); balandin@seas.ucla.edu (A.A.B.)

- **INTRODUCTION**

Research interest in low-dimensional magnetic systems has been rapidly increasing since the experimental discovery of various forms of magnetism in the few- and single-layer limits of quasi-two-dimensional (quasi-2D) materials.[1–5] These developments in materials have enabled designs of novel spintronic devices with applications in sensing, data storage, and information processing.[6–8] In the realm of magnetic materials used in spintronics, there has been a considerable focus on ferro- and ferrimagnetic (FM) compounds that were either electrically insulating or electrically conductive[9,10], while much less attention has been devoted to antiferromagnetic semiconductors (AFMS). Notably, the effect of spin ordering on the dynamics of charge carrier transport and electronic noise in AFMS, especially across the material's magnetic transition, is often overlooked. Exploring the intricate interplay of spins on charge carrier dynamics associated with the defect centers, *i.e.*, spin-dependent trap-assisted generation-recombination (G-R) processes, is crucial for designing novel spintronic devices with more advanced and efficient capabilities.

Transition-metal phospho-trichalcogenides denoted as $MPX_3$, where M is a transition metal and X is a chalcogen, are a subclass of 2D van der Waals (vdW) materials that encompass many AFMS with diverse electrical, optical, and magnetic properties.[11–18] Owing to their low cleavage energy, often even less than that of graphite, these materials can be conveniently exfoliated down to a few or single layers, making them a versatile platform for investigating the physics of low-dimensional materials with potential spintronic device applications.[19–22] The band gaps of these materials span a wide range, extending from 1.3 to 3.5 eV.[11,19] Many members of $MPX_3$ group of materials exhibit various forms of antiferromagnetic (AFM) spin ordering both in bulk and even at the single-atomic layer limit.[22–24] The 'M' element determines the magnetic base state of these $MPX_3$ materials. While $FePS_3$ undergoes an Ising-type transition from AFM to paramagnetic (PM) spin ordering at 118 K, $NiPS_3$ and $MnPS_3$ exhibit an XY-type and Heisenberg transitions at 155 K and 78 K, respectively.[22,25,26] These unique magnetic and electric properties, coupled with the feasibility of their exfoliation down to single atomic layers, position $MPX_3$ as a promising candidate for emerging spintronic applications in ultimate low-dimensional limits.[27–31]

New material systems often require innovative approaches in materials characterization. While obtaining quasi-2D $MPX_3$ magnetic samples is relatively straightforward owing to their low cleavage energy, characterizing magnetism in 2D limits remains a challenging task. Optical techniques such as Raman spectroscopy, photoluminescence, and magnetic atomic force microscopy are often the techniques of choice owing to their high spatial resolution when it comes to studying magnetism in 2D limits.[22,32,33] Other advanced non-optical techniques such as superconducting quantum interference devices and vibrating sample magnetometers, have also been widely used to characterize low-dimensional magnetic materials.[34,35] Low-frequency noise (LFN) spectroscopy has been extensively employed to study the charge carrier dynamics of ferro- and ferrimagnetic materials, but the electronic noise data on AFM materials is scarce.[36,37] By itself, LFN is considered a limiting factor for the magnetic device performance, for example, in magnetic random access memories.[38] Understanding the nature of LFN in a given material is essential for assessing the material's quality and applicability for different electrical, spintronic, or sensor uses.[39–43] On the other side, the noise spectral density behavior under external stimuli such as electrical bias and temperature can provide insights into the characteristics of charge transportation, carrier recombination mechanisms, and even detect magnetic transition.[18,44–48]

The LFN includes the $1/f$, flicker ($f$ is the frequency), generation-recombination (G–R), and random telegraph signal (RTS) noises with both G–R and RTS noise showing Lorentzian-type spectrums.[49–51] These noise types are associated with material defects acting as trap centers for the charge carriers, providing information on charge carrier dynamics and defects.[45,46,52,53] In many ferromagnetic materials, Nyquist and excess noise are among the dominant noise sources.[36] The noise data on magnetic materials reveal the appearance of non-Gaussian RTS noise from thermally generated microscopic processes.[54,55] RTS noise involves the dominance of a single fluctuator of either electrical or magnetic origin that switches between two states.[56,57] The appearance of RTS noise has also been attributed to magnetic domain fluctuations and domain-domain interactions.[58,59] Among a few reported studies of LFM in AFM materials, it was found that thermally agitated magnetic noise in Co-based amorphous and polycrystalline alloys is of $1/f$ type.[60] Note that $1/f$ and RTS noises are dominant noise sources in magnetic sensors and magnetic tunnel junctions[61,62], and the noise data is generally used as a metric to measure the sensitivity and detectability of the device under varying magnetic fields.[61]

Here, we report low-frequency electronic noise in FePS$_3$, a van der Waals AFM semiconductor. We use noise measurements as *noise spectroscopy*, a materials characterization technique, to study the charge carrier dynamics affected by spin ordering, trapping states, and interfaces. Owing to the high resistivity of the material, we conducted measurements in the vertical device configuration using *h*-BN capping in the test structures. The Néel temperature of the magnetic phase transition in FePS$_3$ is $T_N$ = 118 K. Previous studies of FePS$_3$ were focused on its optical, magnetic, and phononic properties.[63–67] The electrical transport data for this AFM semiconductor is limited.[18,68] The challenges with the electrical characterization of FePS$_3$ stem from its high room-temperature (RT) resistivity, which further increases at cryogenic temperatures due to its semiconducting nature.[68] To alleviate this challenge, we designed and fabricated vertical *h*-BN/FePS$_3$ heterostructure devices and measured the cross-plane electrical conductivity across the Néel temperature of FePS$_3$. The noise characteristics across the Néel temperature reveal a complex correlation between the electrical and magnetic properties. Our study shed light on the interplay between electrical carrier dynamics and magnetic properties of AFM semiconductors.

- **MATERIAL CHARACTERIZATIONS AND DEVICE PREPARATIONS**

The bulk FePS$_3$ samples used in this study were synthesized via the chemical vapor transport (CVT) technique (2D Semiconductors, USA). Figure 1 (a) presents the crystal structure and spin configuration of FePS$_3$. The crystal has a monoclinic structure with C2/m symmetry in the bulk form. The Fe metal atoms are arranged in a honeycomb hexagon lattice structure in the crystal and enclosed by (P$_2$S$_6$)$^{4-}$ bipyramids. The direction of P dimers is perpendicular to the Fe honeycomb hexagon. The spin direction follows Ising-type AFM spin ordering with Fe$^{2+}$ ions ferromagnetically coupled with the nearest two of the three atoms, whereas the interlayer spin coupling follows AFM ordering.[11] In terms of magnetic spin configuration, FePS$_3$ follows zigzag AFM ordering where adjacent metal atoms in the lattice along the zigzag chain direction exhibit opposite spin orientations.[23] To verify the quality of the material, we conducted Raman spectroscopy measurements on the bulk crystal. Figure 1(b) provides the Raman spectrum of the bulk FePS$_3$ sample. The obtained Raman peaks are consistent with the published data.[22,69] Figure 1 (c) presents the schematic of the fabricated *h*–BN/FePS$_3$ two-terminal vertical devices with metal contacts at the top and bottom of the active layer. The *h*–BN layer overlapping the FePS$_3$ flake edges prevents any possible lateral connections between the top and bottom contacts, ensuring a vertical charge transport for electrical measurements. Details of our fabrication procedure are provided in the Methods. Figure 1 (d) shows an optical microscopy image of a representative fabricated FePS$_3$ device used for electrical and noise measurements.

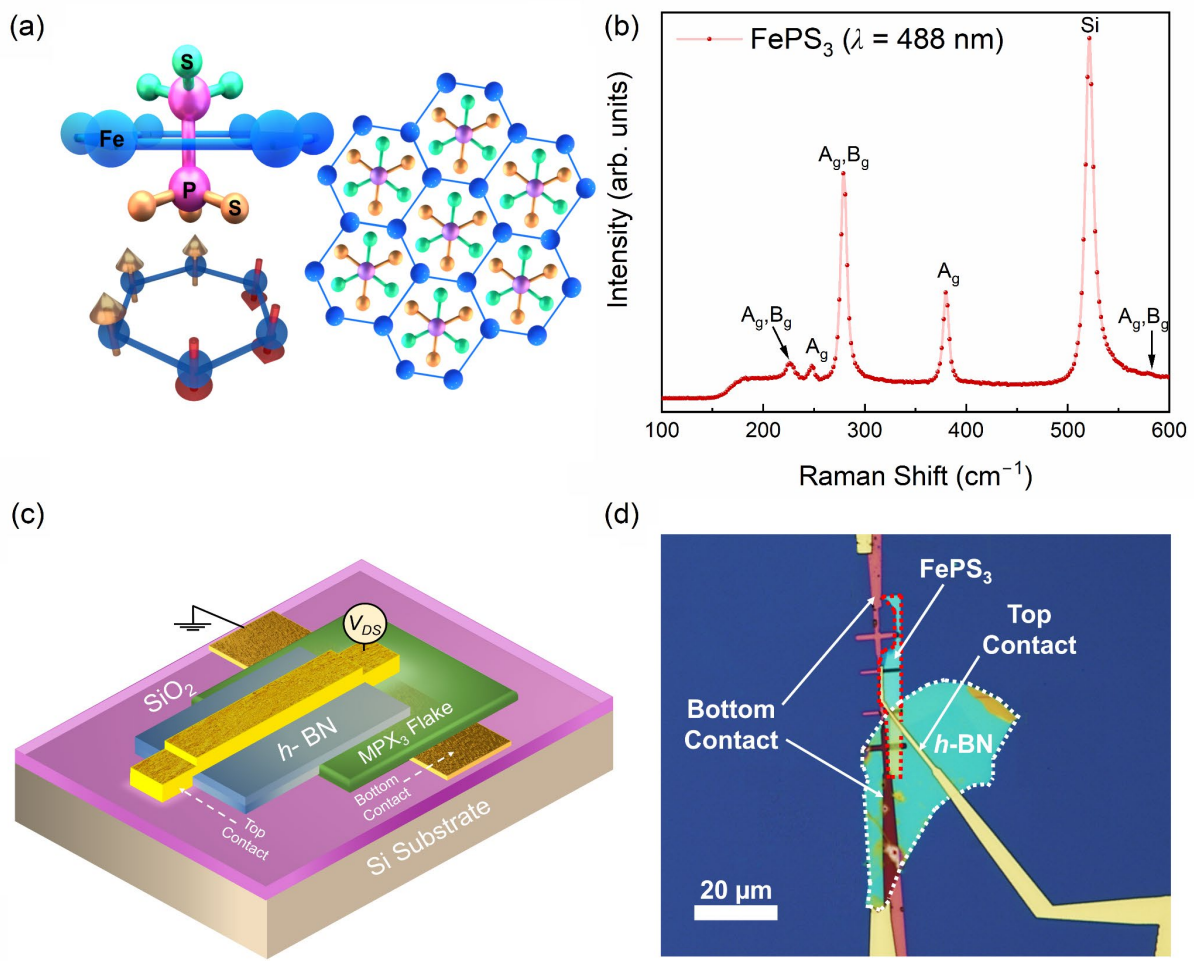

**Figure 1**: (a) Schematic of the crystal structure of FePS$_3$. The Fe atoms are arranged in honeycomb hexagon lattice structures with Ising-type antiferromagnetic spin ordering enclosed by $(P_2S_6)^{4-}$ bipyramids. (b) Room temperature Raman spectrum of an exfoliated FePS$_3$ layer measured under $\lambda$ = 488 nm laser excitation. (c) Schematic of the fabricated $h$-BN/FePS$_3$ vertical heterostructure devices. The $h$–BN layer prevents the direct contact between the top contact and the side of the FePS$_3$ layer, ensuring no lateral charge transport to the bottom contact. (d) Optical microscopy image of a representative FePS$_3$ vertical device.

5 | P a g e

- **EXPERIMENTAL RESULTS AND DISCUSSIONS**

In Figure 2 (a), we present the I-V characteristics of a representative vertical device measured at different temperatures. The data show a linear trend at all examined temperatures, confirming that the fabricated contacts are Ohmic. The measured current levels are significantly higher in the vertical device configuration compared to that for lateral devices.[18,64] The latter is because, in the vertical device configuration, we have larger cross-sections of the channels and much shorter channel lengths, typically in the range of a few tens of nanometers. The latter is substantially lower than the typical dimensions in lateral devices fabricated by conventional lithography methods. The higher current levels are essential for accurate noise measurements. The noise measurement, conducted in a home-built system, confirms the quality of the Ohmic contacts. Figure 2 (b) shows the current spectral density, $S_I$, at frequency $f = 10$ Hz, as a function of device current, $I_{DS}$. The noise data, measured at room temperature, indicate quadratic, $S_I \sim I_{DS}^2$, dependence, which is characteristic of linear resistors. The quadratic dependence confirms that the contacts are not subjected to electron migration-related damage at current levels used in the measurements. More details of our noise experimental setup and measurement procedures can be found in the Methods section and our prior reports for other material systems.[45,70,71]

The channel resistance, $R$, of the $FePS_3$ vertical device as a function of temperature was calculated from the I-V data and is shown in Figure 2 (c). The resistance decreases with the temperature rise, confirming the semiconducting nature of the material. No abrupt changes in $R$ values are observed in the vicinity of the Néel temperature, $T_N$=118 K. To analyze the change in the temperature-dependent resistance of the device near the magnetic transition, we calculated the differential resistance, $dR/dT$, and plotted it as a function of $T$ (see Figure 2 (d)). One can notice that $dR/dT$ attains its extreme value at ~115 K, close to $T_N$, suggesting that the charge transport is affected by the magnetic transition. The anomaly observed in the charge transport in the vicinity of the magnetic transition is attributed to a complex scattering process of charge carriers by phonons and magnons in $FePS_3$ and $MnPS_3$.[72,73] One may expect that the signatures of such processes should be even more pronounced in LFN measurements. The latter explains our unconventional choice of characterization technique for this type of material.

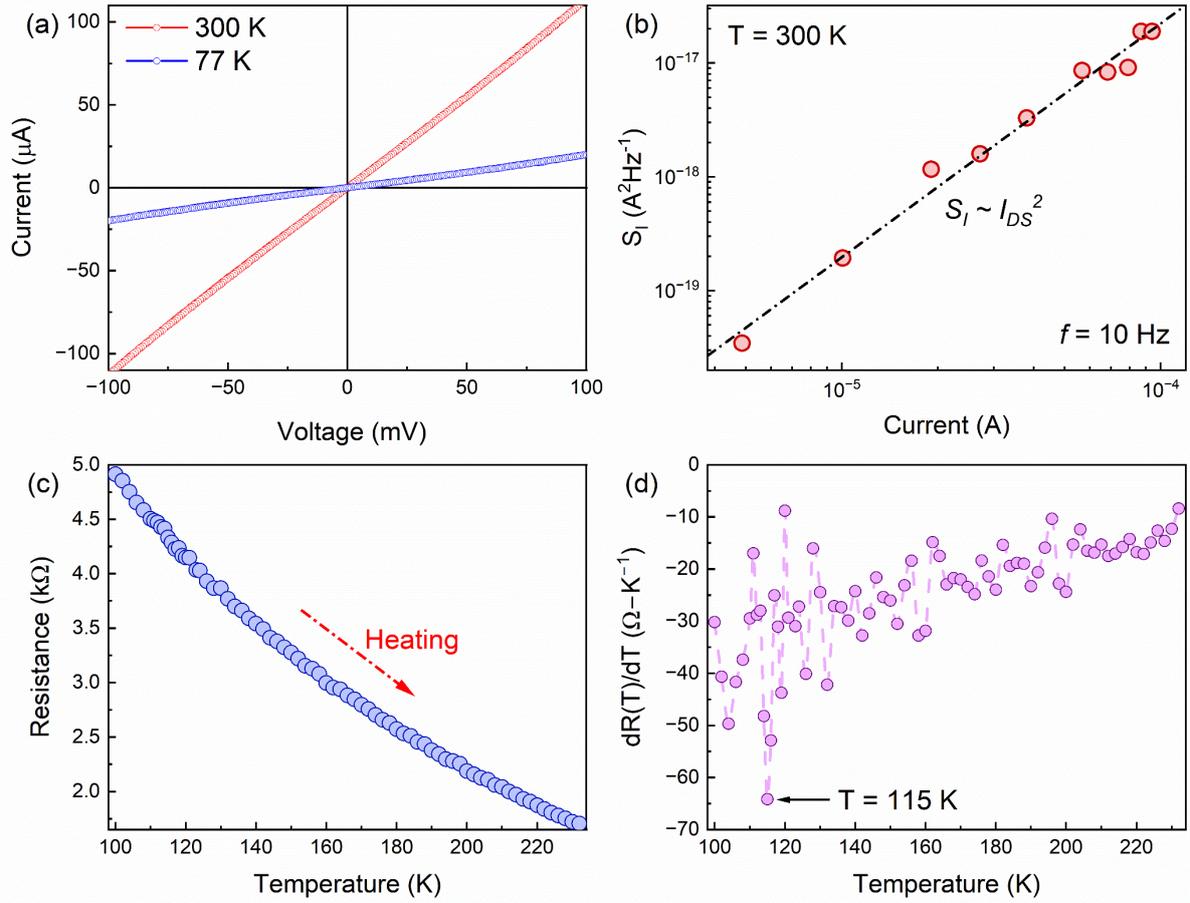

**Figure 2**: (a) Current-voltage characteristics of the *h*-BN/FePS$_3$ vertical device measured at 77 K and 300 K. (b) Noise spectral density, $S_I$ at $f$ =10 Hz plotted as a function of device current, $I_{DS}$. (c) Cross-plane resistance of the FePS$_3$ channel as a function of temperature. (d) The differential resistance, *dR/dT*, as a function of temperature. The data shows the maximum values and fluctuations at ~115 K, close to the Néel temperature of FePS$_3$.

The temperature-dependent LFN spectroscopy measurements were conducted in the temperature range of 100 K to 130 K, across the magnetic phase transition. Figure 3 (a – i) shows the noise current spectral density normalized by current, $S_I/I^2$, as a function of frequency, *f*, measured at different temperatures. In all plots, the source-drain current level was $I_{DS}$ = 10 µA. Overall, the spectra obtained at all temperatures show *1/f* flicker noise type dependence, which is expected for semiconductor materials.[51,74] In addition, two Lorentzian bulges can be seen over the *1/f* flicker noise envelope at different temperatures (also see Supplemental Figure S1 and S2). One Lorentzian bulge, denoted as L1 and depicted by a black solid curve, is present in the spectra over the entire examined temperature range. Another Lorentzian bulge, denoted as L2 and depicted by a red solid line, only appears in the vicinity of the Néel temperature (Figure 3 (d) and 3 (e)). We argue that the origin of these Lorentzian bulges is substantially different, and it presents valuable info for the charge carrier dynamics in this type of material.

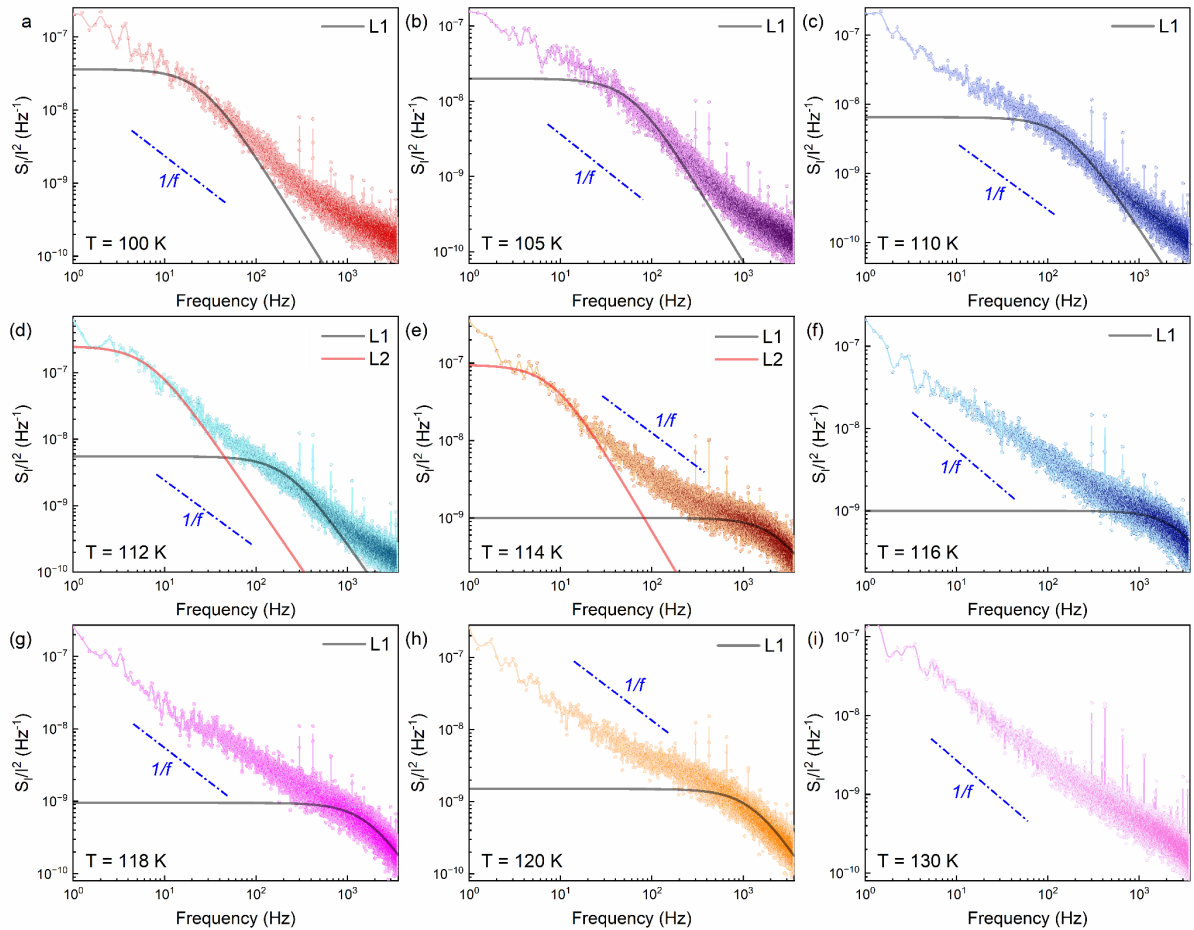

**Figure 3**: Normalized noise spectral density, $S_I/I^2$, of the FePS$_3$ vertical device as a function of frequency, $f$, measured in different temperatures, ranging from 100 K to 130 K. The noise spectra show $1/f$ background with pronounced Lorentzian bulges. The Lorentzian bulge, L1, depicted by the black solid lines, is present at all examined temperatures, and it is attributed to the trap-assisted generation-recombination processes. The second Lorentzian bulge, L2, depicted by the red solid lines, only emerges in the temperature range of 112 K<T<118, close to the Néel temperature of FePS$_3$ antiferromagnetic semiconductor.

To better understand the origin of the Lorentzian components and their temperature dependence, we calculated the normalized spectral density, multiplied by the frequency, $f \times S_I/I^2$, and plotted the data as a function of $f$ for different temperatures. The results are shown in Figure 4 (a, b). The $f \times S_I/I^2$ plot separates the Lorentzian component from the $1/f$ background. The plots reveal each Lorentzian bulge as an individual peak. The maximum value of the peak corresponds to the corner frequency, $f_c$, of the Lorentzian bulges. The pronounced peaks in Figure 4 (a, b), denoted with black solid lines, are associated with the L1 Lorentzian bulges shown in Figure 3. Apart from the black solid lines, at $T=112$ K, the second peak representing L2 emerges at lower frequencies denoted by red solid lines. Both Lorentzian peaks are present in the noise spectra up to $T=118$ K. However, only the $f$ tail of L2 is visible in the temperature range from 115 K to 118 K, with its peak covered by L1. We extracted $f_c$ values of L2 by extrapolating the fitting and finding the maximum values of the fitted curves. Note that the L2 peak disappears at $T = 119$ K as the material crosses its Néel temperature and enters the paramagnetic phase. On the contrary, the L1 peak is seen in the noise spectra in the entire measured temperature range from 100 K to 125 K; at the higher temperature the peak moves beyond the maximum measured frequency (refer to Supplemental Figure S3 (b)). The continuous presence of the L1 peak in the spectra in the entire examined temperature range suggests that this peak is of G-R nature and is associated with a dominant charge carrier trap center due to a specific defect. In contrast, the presence of the L2 peak, only in the temperature range of $112 \text{ K} \leq T \leq 118 \text{ K}$, close to the Néel temperature, ties its origin solely to the material's magnetic phase transition. The Lorentzian features are frequently observed in materials undergoing phase transitions, as discussed below.

To track the temperature characteristics of both peaks, we plotted the corner frequency, $f_c$, as a function of temperature. The results are shown in Figure 4 (c) and Supplemental Figure S4 (a) for a different current level. The $f_c$ value of L1 increases linearly with increasing temperature, with a strong deviation from this monotonic dependence only at temperatures close to the magnetic transition. The $f_c$ of the L2 peak shows an entirely random dependence on temperature, and it is only visible between $T = 112$ K and 118 K. The drastically different temperature-dependent behavior of the two Lorentzian bulges confirms their distinct physical origins. Since the L1 peak remains visible in both AFM and PM phases, we attribute it to G–R noise from a dominant trap. However, it is intriguing that the corner frequency of the L1 peak, and hence its characteristic time constant, $\tau = 1/2\pi f_c$, associated with the G–R process, also depends on the spin reconfiguration across the Neel temperature. The $\tau$ is associated with the G–R trapping and de-trapping and expressed as the combination of capture ($\tau_c$) and emission ($\tau_e$) time constants such as $\frac{1}{\tau} = \frac{1}{\tau_c} + \frac{1}{\tau_e}$ where $\tau_c = 1/\sigma v_T p$ and $\tau_e = 1/\sigma v_T N_v \exp(-E_a/k_B T)$, in which $\sigma$ is the capture cross-section of the G–R trap, $v_T$ is the thermal velocity of the carrier, $N_v$ is the carrier density of states in the conduction band, and $p$ is the equilibrium hole carrier concentration.[75] The $\tau$ value decreases with increasing ambient temperature except near the transition temperature, where it is affected by the material magnetic properties as shown in Figure 4 (d) (also see Supplemental Figures S4 (c) for different device currents). Such a linear decrease in trap assisted characteristic time constant is typical for semiconductor materials and devices.[76,77] However, an anomaly across the magnetic transition suggests that the behavior is not related strictly to the semiconducting nature but to the material's magnetic property, which influences the G-R rate as it passes through the Néel temperature. Studies on magnetic materials confirm that the recombination rate has a strong spin dependency, affecting the material's transport properties.[78] In the case of semiconductors, spin polarization can affect carrier lifetime.[78–80] Therefore, the non-monotonic dependence of corner frequency, $f_c$, and time-constant, $\tau$ are likely associated with the change in trap-assisted carrier G-R rate due to changes in the spin order at the onset of magnetic transition.

Using the temperature-dependent data for $f_c$, one can extract the trap activation energy, $E_a$, following the Arrhenius equation, $f_c = f_0 \exp\left(-E_a/k_B T\right)$, where $f_0$ is the attempt frequency factor, and $k_B$ is the Boltzmann constant.[81] Figure 4 (e) presents the $\ln(f_c)$ vs. $1/T$ plot of the data for the L1 peak. A linear regression, shown with the dashed lines, over the whole data set excluding those in the vicinity of the Neel temperature, yields an activation energy of 0.2 eV. The deviation of the $\ln(f_c)$ in the proximity of magnetic transition from the linear trend affirms that the activation energy varies with spin reconfiguration as well. It has been found that a strong interplay between magnetic order and transport properties can change the carrier activation energy at the Néel temperature.[82] Studies on rare-earth ferromagnetic materials reveal a similar non-trivial $f_c$ and $E_a$ dependence across the magnetic phase transition governed by thermally activated discrete fluctuator due to coupling between magnetic and resistance fluctuations.[81,83]

The origin of the L2 peak is relatively straightforward. Since it appears only close to the magnetic phase transition at ~118 K and disappears in the PM phase, it is evident that the Lorentzian component is associated with the magnetic phase transition and subsequent subtle change in its resistive behavior, as shown in Figure 2(d). The nature of noise spectra is generally affected by the change in material phases in terms of electrical, structural, and magnetic properties.[62,84,85] When the material undergoes a phase change, the noise behavior can appear in the Lorentzian form due to the creation of a two-level system. In our case, at 118 K, the material transitions from PM to AFM phase. At the transition point, the material system can fluctuate between the two phases, acting as a two-level system. The magnetic phase transition is observed in many magnetic materials, characterized by a two-level resistance state.[84–86] The two-level system can also be related to the discrete fluctuations associated with the RTS noise.[81] The RTS noise appears in the time domain as pulses with uniform amplitude but random pulse widths and intervals. Such noise often appears during the magnetic phase change.[87] However, our time-domain analysis does not reveal clear signatures of RTS noise across the phase transition as shown in Supplemental Figure S5. This can be attributed to the fact that the phase transition within the system does not occur simultaneously across the entire sample, and the system behaves as it has several two-level system fluctuators.[48]

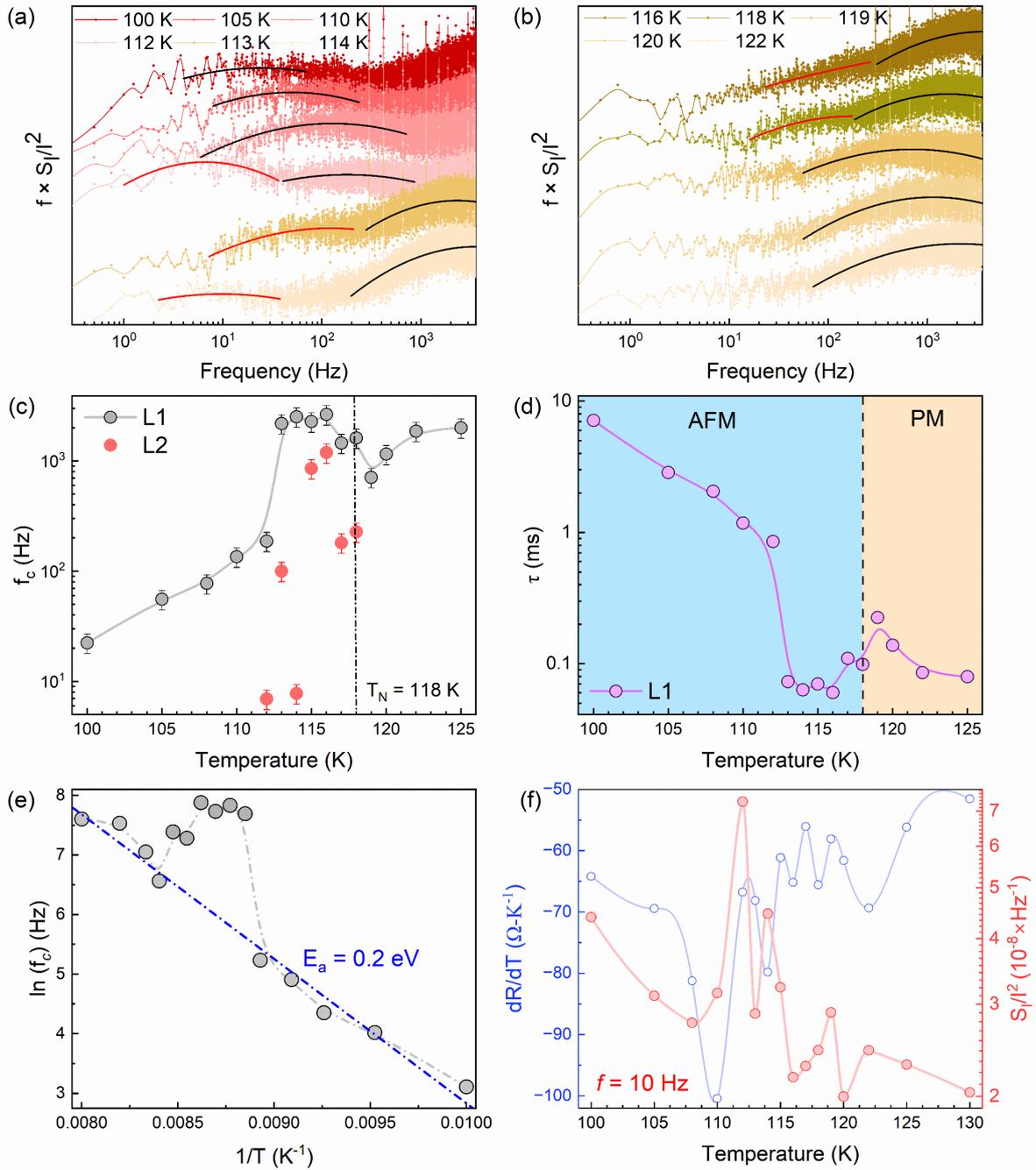

**Figure 4**: (a, b) Normalized noise spectral density multiplied by the frequency, $f \times S_I/I^2$, as a function of frequency and temperature. (c) The corner frequency, $f_c$, of the Lorentzian bulges shown in panels (a, b) plotted as a function of temperature. The error bars represent a 20% margin of error for the extracted data. (d) The trapping time constant, $\tau$, as a function of temperature. (e) The $ln(f_c)$ as a function of inverse temperature, $1/T$. All data points, except those close to the magnetic transition, are fitted by a linear regression. The slope of the line represents the average trap activation energy. (f) The normalized noise spectral density, $S_I/I^2$, measured at $f = 10$ Hz and $I_{DS} = 10$ µA plotted across the Neel temperature of the FePS$_3$ device and compared with the resistance derivative $dR/dT$. The $dR/dT$ minimum and the maximum noise level coincide.

LFN is extremely sensitive to different types of phase transitions.[18,47,88] This characteristic introduces the LFN as a reliable tool to detect the magnetic transition temperature with high accuracy and sensitivity. Figure 4 (f) presents the normalized noise spectral density, $S_I/I^2$, measured at $f = 10$ Hz between 100 K and 130 K at the current $I_{DS} = 10$ μA and the corresponding $dR/dT$ measured during the noise measurements. The noise level experiences a sharp increase near the Néel temperature, followed by several low-intensity peaks (also see Supplemental Figure S6). The most prominent peak occurs at the maximum $dR/dT$ value. The correlation between the highest noise level across the resistive transition has been observed in many materials.[47,70] Specifically for magnetic metals, it has been observed that both resistance and magnetic fluctuations co-exist across the transitions.[89–91] In the case of Cr, which is an AFM transition metal, a two-order increase in the noise level across its Neel temperature has been reported.[92] The latter is attributed to the dynamics of transverse spin density waves.

- **CONCLUSIONS**

We investigated low-frequency current fluctuations in FePS$_3$ van der Waals layered AFMS. Owing to the high resistivity of the material, we conducted measurements on vertical device configuration. The measured noise spectra reveal pronounced Lorentzian peaks of two different origins. One peak is observed only near the Néel temperature, and it is attributed to the corresponding magnetic phase transition. The second Lorentzian peak, visible in the entire measured temperature range, has the characteristics of the G-R processes similar to those in conventional semiconductors but shows a clear effect of the spin order reconfiguration near the Néel temperature. The obtained results contribute to understanding the electron and spin dynamics in this type of AFMS and demonstrate the potential of electronic noise spectroscopy for advanced materials characterization.

- **METHODS**

**Device Fabrication:** To fabricate the device, at first, the bottom metal contacts with a thickness of ~30 nm (Ti/Au) were prepared on top of a clean Si/SiO$_2$ substrate (University Wafer, $p$-type Si/SiO$_2$, <100>) using electron-beam lithography (EBL) for contact patterning and electron-beam evaporation (EBE) system for metal deposition. The bulk FePS$_3$ samples were mechanically exfoliated on top of a PDMS sheet, placed on a glass stamp, and transferred on top of the bottom metal contacts with the help of a dry sample transfer system. An insulating $h$-BN exfoliated thin layer was transferred on top of the FePS$_3$ flakes, partially covering it. In the end, the top contact was patterned over the $h$–BN/FePS$_3$ heterostructure using the EBL technique, and the metals (Ti/Au, 15/150 nm) were deposited with the EBE system.

**Electrical Conductivity and Noise Measurements:** The temperature-dependent electrical measurements were conducted inside a cryogenic probe station chamber (Lakeshore TTPX) between 100 K–200 K under high vacuum (~5×10$^{-6}$ torr). A semiconductor analyzer (Agilent b1500a) was used to measure the current, $I_{DS}$ across the vertical channels under a constant voltage ($V_{DS}$ = 10 mV) in a 2-terminal configuration in an automatic measurement setup at varying temperatures (Lakeshore Model 336 temperature controller). The noise system consists of a noise circuit, a preamplifier (SR 560), and a dynamic signal analyzer (Photon+). The noise circuit contains the device under test (DUT), $R_D$, and a load resistor, $R_L$, in series to a low-noise DC battery. A potentiometer (POT) is also connected to control the voltage drop across the circuit. During the noise measurements, the output voltage fluctuation ($\Delta V$) is transferred to the preamplifier, which amplifies the input signal. The amplified output voltage from the preamplifier is transferred to the signal analyzer, which converts the time-domain signal to its corresponding frequency-varying voltage spectral density, $S_V$. The obtained voltage spectral density data is converted to its equivalent short circuit current spectral density, $S_I$ vs. $f$, where $f$ is frequency.

**Acknowledgment**

FK and ABB acknowledge funding from the National Science Foundation (NSF), Division of Material Research (DMR) *via* project No. 2205973 entitled "Controlling Electron, Magnon, and Phonon States in Quasi-2D Antiferromagnetic Semiconductors for Enabling Novel Device Functionalities."

**Conflict of Interest**

The authors declare no conflict of interest.

**Author Contributions**

AAB and FK coordinated the project and led the data analysis and manuscript preparation; SG fabricated vertical devices, conducted current-voltage and noise measurements, and contributed to the data analysis; ZEN prepared vertical heterostructure samples. All authors contributed to the manuscript preparation.

**The Data Availability Statement**

The data that support the findings of this study are available from the corresponding author upon reasonable request.